\documentclass[aps,prl,amsmath,amssymb,twocolumn,superscriptaddress,showpacs]{revtex4}

\usepackage{amsmath}
\usepackage{graphics}
\usepackage{natbib}

\begin{document}

%\author{Konstantin Y. Bliokh}

%\affiliation{Applied Optics Group, School of Physics, National University of Ireland, Galway, Galway, Ireland}

%\pacs{42.25.Dd, 42.65.Pc, 05.60.Gg} \maketitle

\begin{flushleft}
\large
\textbf{Comment on ``Optical orbital angular momentum from the curl of polarization''}
\end{flushleft}

\normalsize

%\freeline

Recently, Wang \textit{et al.} reported prediction and observation of the ``\textit{new category of optical orbital angular momentum} (OAM)'' [1]. It is known that the angular momentum (AM) of light is divided into the spin angular momentum (SAM), associated with the polarization helicity and the OAM, associated with the azimuthal phase gradient [2]. We argue that the AM described in [1] is not a new OAM, but rather represents the \textit{well-known SAM of light}. Moreover, \textit{paraxial theory} used by Wang \textit{et al.} cannot adequately describe their experiment with \textit{tightly focused field}. In our opinion, the orbital motion of the particles observed in [1] is caused by the OAM generated as a result of \textit{spin-to-orbit AM conversion} upon focusing by a high-NA objective [3].

The linear momentum density (or energy flow density) of an optical field is proportional to the time-averaged Poynting vector, which for monochromatic complex electric field ${\bf E}\left( {\bf r} \right)e^{ - i\omega t}$ reads ${\bf P} \propto {\mathop{\rm Im}\nolimits} \left[ {{\bf E}^* \times \left( {\nabla  \times {\bf E}} \right)} \right]$. In the first paraxial approximation, taking into account small $z$-components of the field, one has
\begin{equation}
\label{field}
{\bf E} = \left[ {{\bf E}_ \bot   + {i k^{-1}}{\bf \hat e}_z \left( {\nabla  \cdot {\bf E}_ \bot  } \right)} \right]e^{ikz}~. 
\end{equation}
Here ${\bf E}_ \bot   = A\left( {x,y} \right)\left[ {\alpha \left( {x,y} \right){\bf \hat e}_x  + \beta \left( {x,y} \right){\bf \hat e}_y } \right]$ is the transverse electric field, where the complex amplitude $A = ue^{i\psi }$ and normalisation $\left| \alpha  \right|^2  + \left| \beta  \right|^2  = 1$ are assumed [1]. It was examined in details recently [4--8] that the momentum density can be divided into orbital and spin parts: ${\bf P} = {\bf P}^O  + {\bf P}^S$, which in the paraxial approximation yield
\begin{equation}
\label{flows}
{\bf P}^O  \propto {\mathop{\rm Im}\nolimits} \left[ {{\bf E}^*  \cdot \left( \nabla  \right){\bf E}} \right]~,~~
{\bf P}^S  \propto \nabla  \times {\mathop{\rm Im}\nolimits} \left( {{\bf E}^*  \times {\bf E}} \right)/2~.
\end{equation}
Using Eq.~(1) and neglecting corrections from $E_z\propto k^{-1}$, we derive the transverse momentum densities (2):
\begin{subequations}
\begin{align}
\label{3a}
{\bf P}_ \bot ^O & \propto u^2 \nabla \psi  + u^2 {\mathop{\rm Im}\nolimits} \left[ {\alpha ^* \nabla \alpha  + \beta ^* \nabla \beta } \right]~,
\\
\label{3b}
{\bf P}_ \bot ^S & \propto \nabla  \times \left( {u^2 \sigma \,{\bf \hat e}_z } \right)/2~,
\end{align}
\end{subequations}
where $\sigma = 2{\mathop{\rm Im}\nolimits} \left( {\alpha^* \beta } \right)$ is the polarization helicity. The two summands of Eq.~(3a) and Eq.~(3b) correspond to the terms ${\bf P}_ \bot ^{(1),(2)}$ and ${\bf P}_ \bot ^{(3)}$, Eq.~(1), in [1]. According to [2,4--8] the OAM and SAM densities are given by ${\bf L} = {\bf r} \times {\bf P}^O$ and ${\bf S} = {\bf r} \times {\bf P}^S$, and using Eqs.~(3), we obtain
\begin{subequations}
\begin{align}
\label{4a}
L_z  & \propto u^2 \partial _\phi  \psi  + u^2 {\mathop{\rm Im}\nolimits} \left[ {\alpha ^* \partial _\phi  \alpha  + \beta ^* \partial _\phi  \beta } \right]~,
\\
\label{4b}
S_z & \propto r \, \partial _r \left( {u^2 \sigma } \right)/2~.
\end{align}
\end{subequations}
Evidently, the terms $J_z^{(1),(2),(3)}$, Eq.~(2) in [1], which were interpreted as OAM, correspond to the two terms of the OAM, Eq.~(4a), and the SAM, Eq.~(4b). Thus, the ``new category of the OAM'' described by $J_z^{(3)}$ is nothing but the SAM of light. Wang \textit{et al.} considered an example with nearly uniform intensity where the SAM originates exclusively from the helicity gradient $\partial _r \sigma$, but in the general case it arises from both $\sigma$ and $u^2$ gradients [4].

In the experiment [1], a paraxial state $\bf{E}\simeq \bf{E}_{\perp}$ with nonuniform polarization $\partial _r \sigma \neq 0$ was prepared,  which carries finite $S_z$ and $L_z =0$ (because $\bf{E}_{\perp}$ was $\phi$-independent). After that, the field was tightly focused by high-NA objective (NA=0.7), ${\bf E}\rightarrow {\bf E}^{f}$, and apparently the AM contributions were calculated from paraxial Eqs.~(4) for the resulting field ${\bf E}^{f}_{\perp}$ (Fig. 3b in [1]). This is erroneous since the focused field is \textit{significantly nonparaxial}, and the longitudinal component $E_{z}^{f}\neq 0$ must be taken into account. Apparently, $J_z^{(3)}$ in Fig.~3b [1] is a part of the SAM of the nonparaxial field. At the same time, tight focusing is known [3,8-10] to produce spin-to-orbit AM conversion. It generates non-zero OAM $L_{z}^{f}\propto E_{z}^{f*} \partial _\phi  E_{z}^{f}$, because the $E_{z}^{f}$ component contains vortex $e^{i\sigma\phi}$ for the circularly polarized ($\sigma=\pm 1$) fields even if ${\bf E}_{\perp}$ was $\phi$-independent [3,8-10]. In the postparaxial approximation this OAM can be estimated as $\sim \theta_{0}^{2}/4$ [9], where $\theta_{0}$ is the aperture angle. Hence, the spin-to-orbit conversion is about 10\% for the aperture angles  $\theta_{0}\sim 30^\circ \div 40^\circ$, which is sufficient to cause the orbital motion of particles observed in [1]. 

Finally, we remark that mechanical action of both the spin and orbital energy flows on particles crucially depend on the particle properties [5,8]. Particles used in [1] are rather large compared to the typical scale of the AM density variations, and the assumption of the the local action of the momentum density cannot be justified.
 
\normalsize
\begin{flushleft}
Aleksandr Y. Bekshaev
\small

~~~I. I. Mechnikov National University, Odessa, Ukraine

\end{flushleft}
\normalsize

\begin{flushleft}
Konstantin Y. Bliokh
\small

~~~Applied Optics Group, School of Physics, 

~~~National University of Ireland, Galway, Ireland
\end{flushleft}
\normalsize

\begin{flushleft}
Marat S. Soskin
\small

~~~Institute of Physics, 

~~~National Academy of Science of Ukraine, Kiev, Ukraine
\end{flushleft}

\begin{flushleft}
\end{flushleft}


\normalsize


\begin{thebibliography}{99}


\bibitem{Wang} X.-L. Wang \textit{et al.}, Phys. Rev. Lett. \textbf{105}, 253602 (2010).

\bibitem{OAM} L. Allen \textit{et al.}, Prog. Opt. \textbf{39}, 291 (1999).

\bibitem{Zhao} Y. Zhao \textit{et al.}, Phys. Rev. Lett. \textbf{99}, 073901 (2007).

\bibitem{BS} A. Y. Bekshaev and M. S. Soskin, Opt Commun. \textbf{271}, 332 (2007).

\bibitem{Berry} M. V. Berry, J. Opt. A: Pure Appl. Opt. \textbf{11}, 094001 (2009).

\bibitem{Li} C.-F. Li, Phys. Rev. A \textbf{80}, 063814 (2009).

\bibitem{Bliokh} K. Y. Bliokh \textit{et al.}, Phys. Rev. A \textbf{82}, 063825 (2010).

\bibitem{BBS} A. Y. Bekshaev \textit{et al.}, arXiv:1011.0862.

\bibitem{Nieminen} T. A. Nieminen \textit{et al.}, J. Opt. A: Pure Appl. Opt. \textbf{10}, 115005 (2008).

\bibitem{Oscar} O. G. Rodr\'{i}guez-Herrera \textit{et al.}, Phys. Rev. Lett. \textbf{104}, 253601 (2010).

\end{thebibliography}
\end{document}